\documentclass[preprint]{ptephy_v1}

\preprintnumber{2305.19388} 
\usepackage{hyperref}
\usepackage{orcidlink}

\begin{document}

\title{A closer look at the  Yukawa's interaction from a symmetry group perspective}


\author{Luiz L. Lopes}
\affil{ Centro Federal de Educa\c{c}\~ao Tecnol\'ogica de Minas Gerais Campus VIII, Varginha/MG, CEP 37.022-560, Brazil \email{llopes@cefetmg.br}
}%


\begin{abstract}%
I investigate the use of the SU(3) Clebsch-Gordan coefficients in light of the relations of completeness and closure. I show  that in the case of $\alpha_V = F/(F+D)~\neq$ 1, there is an additional interaction: the exchange of a $\rho$ meson between a $\Lambda$ and a $\Sigma^0$ hyperon that only affects the symmetric coupling. I then calculate these additional coupling constants and show that this recovers the completeness and closure of the SU(3) Clebsch-Gordan coefficients for all values of $\alpha_V$. Besides, it increases the symmetry of the theory, once now we can group the baryon octet into four doublets. Finally, I add the new coupling constants to study numerical results in the hyperon onset in dense nuclear matter  assuming $\alpha_V$ as a free parameter.
\end{abstract}

\subjectindex{D01, D06, D41, E32}

\maketitle

\section{Introduction}

The study of nuclear physics is almost a century old. And despite its senility, some techniques developed in the early years are still helpful today in describing strongly interacting matter. In 1935, H. Yukawa~\cite{Yukawa} proposed that the interaction between nucleons was mediated by an exchange of massive particles. Nowadays, such interaction is called a one-boson exchange, or Yukawa coupling~\cite{Bethe1951}, and it is expressed as the so-called Yukawa Lagrangian:

\begin{equation}
    \mathcal{L}_{YUK} = -g_{BBM}(\bar{\psi}_B\psi_B)M. \label{YukLag}
    \end{equation}

The theory of strong force and the use of the  Yukawa couplings had a great leap with the works of J. Schwinger~\cite{SCHWINGER1957} and especially with the elegant and imperative work of J. J. Sakurai~\cite{SAKURAI1960}.  Based on current conservations and local gauge invariance, Sakurai proposed a model that deals explicitly with baryon-baryon interaction via vector mesons exchange. In such a model, the $\omega$ meson couples to the hypercharge while the $\rho^0$ meson couples to the isospin.

With the development of symmetry group theories, Sakurai's theory was relegated  as just a particular case of the  more powerful and well-accepted flavor SU(3) symmetry group theory~\cite{Gell1962,Beh1962, Swart1963,DOVER1984}. However, with the onset of the more restrictive flavor-spin hybrid SU(6) group: SU(6) $\supset$ SU(3) $\otimes$ SU(2), Sakurai's theory was restored in its full glory; and  again, the $\omega$ meson couples to the hypercharge and the $\rho$ meson couples to the isospin~\cite{DOVER1984, Pais1964,Pais1966,NAGELS1979}. 

Although the Yukawa coupling explicitly deals  with baryon-baryon interaction via one-boson exchange, such interaction has proven extremely useful also in many-body theories. In 1974, J. D. Walecka applied the Yuakwa coupling to describe dense nuclear matter in mean field approximation (MFA)~\cite{WALECKA1974}. In this approach, the mesonic fields are replaced by their expected values and the nucleons do not interact with each other but instead, they behave like a free Fermi gas with a classical background field.
The Walecka model and its extensions are today known as quantum hadrodynamics~\cite{Serot_1992} and soon become a standard effective field theory to describe dense nuclear matter.

From the early 1990s on, the interest in studying neutron stars with exotic matter has increased significantly, and to reduce the huge uncertainties about the hyperon-meson coupling constants, the use of the SU(6) symmetry group became a standard approach and is widely used, even in nowadays~\cite{ELLIS1991,SCHAFFNER1994,Schaffner1996,Pal1999,Schaffner2000,WEISSENBORNNPA,Tolos2017,Lopes2020EPJA}. However, the discovery and confirmation of hypermassive neutron stars in the earlier 2010s have shaken our trust in SU(6) coupling constants. For instance, the J0348+0432 with a mass range of $2.01 \pm 0.04~M_\odot$~\cite{Antoniadis} and especially the PSR J0740+6620, whose gravitational mass is 2.08 $\pm$ 0.07 $M_\odot$~\cite{Miller2021,Riley2021} bring great tension between the  onset of energetically favorable hyperons and its well-known softening of the equation of state (EoS). This phenom is called the hyperon puzzle. Quickly, several authors realized that it was possible to reconcile massive neutron stars with hyperons in their core by partially breaking the SU(6) symmetry in favor of the less restrictive flavor SU(3) symmetry~\cite{WeissPRC2012,WeissPRC2014,Miyatsu2013,Gusakov2014,Micaela2016,Li2018,lopesnpa,Dex2021,Lopes_2022,Lopes2022CTP,HongCPC,mateus,FU2022,lopesPRD,Sedrakian2023}.

Although in the SU(3) the $\rho^0$ meson does not necessarily couples direct to the isospin, its sign depends on the isospin projection~\cite{Swart1963,DOVER1984}. This implies that the coupling of the $\rho$ between the neutrons is the opposite of those between the protons. The same is true for the $\Xi$'s and for the $\Sigma$'s.  Such behavior is summarized in Eq.~\ref{eq2}.

\begin{eqnarray}
g_{nn\rho} = - g_{pp\rho}, \quad g_{\Xi^-\Xi^-\rho} = - g_{\Xi^0\Xi^0\rho}, \nonumber \\
g_{\Sigma^-\Sigma^-\rho} = - g_{\Sigma^+\Sigma^+\rho}. \label{eq2}
\end{eqnarray}

Moreover, as someone can correctly guess, the coupling constant between $\Lambda$'s and between $\Sigma^0$'s are null:

\begin{eqnarray}
g_{\Lambda\Lambda\rho} =  g_{\Sigma^0\Sigma^0\rho} = 0, 
\end{eqnarray}
once their isospin projection is zero. 

When we are dealing with the Yukawa coupling (Eq;~\ref{YukLag}), especially in quantum hadrodynamics, we usually assume that Dirac field $\bar{\psi}_B$ is the complex conjugate of the field $\psi_B$. From the SU(3) point of view, that is almost always true. Most of the $g_{BBM}$ is zero for crossed terms -i.e.; if $\bar{\psi}_B$ and $\psi_B$ are not complex conjugates to each other. 

The KEY point of the present work is that if we assume that $\bar{\psi}_B$ and $\psi_B$ are always complex conjugates to each other, the relation of completeness and closure of the SU(3) Clebsch-Gordan coefficients is violated if $\alpha_V$~$\neq$ 1.  This implies that,  in this case, the set of coupling constants is incomplete. Indeed, there are  crossed Yukawa couplings (sometimes called coupled channels):

\begin{equation}
-g_{\Sigma^0\Lambda\rho}(\bar{\psi}_{\Sigma^0}\psi_\Lambda)\rho^0, \quad \mbox{and} \quad
-g_{\Lambda\Sigma^0\rho}(\bar{\psi}_{\Lambda}\psi_{\Sigma^0})\rho^0, \label{cross}
\end{equation}
that may in fact differ from zero.  From the field theory point of view~\cite{SCHWINGER1957}, the Eq.~\ref{cross} indicates that the $\Sigma^0$ and the $\Lambda$ interact with each other via $\rho$ meson exchange. However, in the mean-field approximation, the $\Lambda$ and the $\Sigma^0$ now interact with the background field of the meson $\rho$. The strength of this interaction depends only on the coupling constant.

In this work, I calculate the crossed coupling constants from Eq.~\ref{cross} by imposing that the Yukawa Lagrangian (Eq.~\ref{YukLag}) is invariant under the SU(3) flavor symmetry group and show that this  crossed coupling contributes to the symmetric coupling while having no effect in the antisymmetric one. Therefore, it restores the relation of completeness and closure for the symmetric coupling, and as a consequence,  for all values of $\alpha_V$. Thereafter, I explicitly add the crossed Yukawa terms to build a more complete QHD Lagrangian. Then, I calculate the new energy eigenvalues for the $\Lambda$ and $\Sigma^0$ hyperons. Finally, we see how the modified energy eigenvalues affect some of the microscopic and macroscopic properties in neutron stars and dense nuclear matter  assuming $\alpha_V$ as a free parameter.

\section{ The SU(3) Group Formalism}

In the SU(3) symmetry group formalism (see ref.~\cite{Swart1963,DOVER1984,Pais1964,Pais1966,lopesPRD} and the references therein to additional discussion), each eigenstate can be labeled as $|N~Y~I~I_3 \rangle$,  where $N$ is the dimension of the representation, $Y$ is the hypercharge, $I$ is the total isospin and $I_3$ is the isospin projection. Assuming that the Yukawa coupling of the QHD (Eq.~\ref{YukLag}) is invariant under the SU(3) flavor symmetry group, implies that its eigenstate is $|0~0~0~0 \rangle$, or simply a unitary singlet. 

The eigenstate of the $\rho^0$ is $|8~0~1~0\rangle$. Therefore, in order to produce a Yukawa Lagrangian that is a unitary singlet, the direct product ($\bar{\psi}_B~\otimes~\psi_B$) also must have the same eigenstate: $|8~0~1~0\rangle$. As the hypercharge and isospin projection are additive numbers, the simplest way to couple ($\bar{\psi}_B~\otimes~\psi_B$)  to result in $|8~0~1~0\rangle$ is to assume that $\bar{\psi}_B$ and $\psi_B$ are complex conjugates to each other. After that, we must couple the resulting   $|8~0~1~0\rangle$ state to the $\rho^0$ meson in order to obtain the unitary singlet: ($\bar{\psi}_B~\otimes$~$\psi_B$) $\otimes~ \rho^0$ =  $|0~0~0~0 \rangle$. 
From the use of the Speiser method~\cite{Swart1963}, there are two ways to couple ($\bar{\psi}_B~\otimes~\psi_B$) to result in the  $|8~0~1~0\rangle$  state, typically, the antisymmetric and the symmetric coupling. Therefore, the Yukawa Lagrangian of Eq.~\ref{YukLag}  can be rewritten as:

\begin{equation}
\mathcal{L}_{\rm Yukawa} = -((gC_8 + g'C'_8 ) \times  C_1) (\bar{\psi}_B\psi_B)\rho^0, \label{expY}
\end{equation}

The $g$ ($g'$) is the constant associated with the symmetric (antisymmetric) coupling, while the  $C_8$ ($C'_8$) is the SU(3) Clebsch-Gordan (CG) coefficients of the symmetric (antisymmetric) coupling to result in the $|8~0~1~0\rangle$ state. Furthermore, $C_1$ is the CG coefficients to the product $(\bar{\psi}_B \psi_B)\times \rho^0$  to result in the unitary singlet.
The SU(3) CG coefficients can be calculated from the isoscalar factors, as discussed in Ref.~\cite{Swart1963}. Once its values are well known, we use the tables presented in Ref.~\cite{CG}. Explicitly, we have:

\begin{align}
g_{pp\rho}={}& -\left( - \sqrt{\frac{3}{20}}g - \sqrt{\frac{1}{12}}g' \right) \times \sqrt{\frac{1}{8}},  \nonumber\\ 
g_{nn\rho}={}& -\left( - \sqrt{\frac{3}{20}}g - \sqrt{\frac{1}{12}}g' \right) \times -\sqrt{\frac{1}{8}}, \nonumber \\ 
g_{\Lambda\Lambda\rho}={}& -\left( 0g + 0g' \right) \times 0,  \nonumber \\
g_{\Sigma^0\Sigma^0\rho}={}& -\left( 0g + 0g' \right) \times 0, \nonumber \\
g_{\Sigma^+\Sigma^+\rho}={}& - \left(0g - \sqrt{\frac{1}{3}}g' \right) \times  \sqrt{\frac{1}{8}}, \nonumber  \\ 
g_{\Sigma^-\Sigma^-\rho}={}& - \left(0g + \sqrt{\frac{1}{3}}g' \right) \times  \sqrt{\frac{1}{8}},  \nonumber \\ 
g_{\Xi^0\Xi^0\rho}={}& -\left( - \sqrt{\frac{3}{20}}g + \sqrt{\frac{1}{12}}g' \right) \times - \sqrt{\frac{1}{8}}, \nonumber \\
g_{\Xi^-\Xi^-\rho}={}& -\left( - \sqrt{\frac{3}{20}}g + \sqrt{\frac{1}{12}}g' \right) \times  \sqrt{\frac{1}{8}}, \label{CGresults}
\end{align}

Nevertheless, the SU(3) CG coefficients, as their SU(2) counterparts (see for instance chapter 3 in Sakurai's classical book~\cite{Sakuraibook}), must satisfy the relations of completeness and closure. In other words, we must have: $\sum C^2_8 = \sum C'^{2}_8 = \sum C^2_8 = 1$. However, one can easily check that:

 \begin{equation}
\left \{
\begin{array}{cc}
   C_8^2  & = 0.6 \\
   C'^2_8  &  =1 \\
   C_1^2   & = 0.75.
\end{array}
\right . \label{closure}
 \end{equation}

The results in Eq.~\ref{closure} show us that the set of coupling constants presented in Eq.~\ref{CGresults} are complete for the antisymmetric coupling ($g'$),  but not complete for the symmetric one ($g$). There are some   additional ($\bar{\psi}_B~\otimes~\psi_B$) product that still results in the
 $|8~0~1~0\rangle$ state, but are not complex conjugate to each other. Indeed, the direct product $\bar{\psi}_{\Sigma^0}~\otimes~\psi_\Lambda$, as well the $\bar{\psi}_{\Lambda}~\otimes~\psi_{\Sigma^0}$  produce an eigenstate $|8~0~1~0\rangle$. The coupling constants $g_{\Sigma\Lambda\rho}$ and $g_{\Lambda\Sigma\rho}$ can be calculated with the  SU(3) Clebsch-Gordan (CG) coefficients:

\begin{eqnarray}
 g_{\Sigma^0\Lambda\rho} =  - \bigg (- \sqrt{\frac{1}{5}}g + 0g' \bigg ) \times \sqrt{\frac{1}{8}} ,\nonumber \\
g_{\Lambda\Sigma^0\rho} =  - \bigg (- \sqrt{\frac{1}{5}}g + 0g' \bigg ) \times \sqrt{\frac{1}{8}}. \label{CG}
\end{eqnarray}

As can be seen, these crossed couplings are only non-null in the symmetric coupling ($g$), as in the antisymmetric one ($g'$),  the set was already complete. When we add these two  additional coupling constants, we recover the relations of completeness and closure:
$\sum C^2_8 = \sum C'^{2}_8 = \sum C^2_8 = 1$, implying that we now have a complete set of coupling constants in agreement with the SU(3) group  for both: the antisymmetric and symmetric couplings. Moreover, as can be seen, unlike the cases of isospin doublets (as protons and neutrons; $\Xi^0$ and $\Xi^-$, etc) the  $g_{\Sigma\Lambda\rho}$ and  $g_{\Lambda\Sigma\rho}$ are both positives and not opposite to each other as the ones in Eq.~\ref{eq2}. Now, following ref.~\cite{Swart1963} we introduce the coupling constants:

\begin{eqnarray}
g_8 = \frac{\sqrt{30}}{40}g + \frac{\sqrt{6}}{24}g', \quad \mbox{and} \quad \alpha_V = 
\frac{\sqrt{6}}{24}\frac{g'}{g8} ,
\end{eqnarray}
which results in:

\begin{eqnarray}
  g_{\Sigma^0\Lambda\rho} =   g_{\Lambda\Sigma^0\rho} = \frac{2}{3}\sqrt{3}g_8(1 - \alpha_V), \quad
  \mbox{implying} \nonumber \\
  \frac{g_{\Sigma^0\Lambda\rho}}{g_{NN\rho}} = \frac{2}{3}\sqrt{3}(1 - \alpha_V).  \label{ncc}    
\end{eqnarray}

Within the flavor SU(3) symmetry, we have in principle
three free parameters: $\alpha_V$, the ratio $z =g_8/g_1$, and the mixing angle $\theta_V$. - see ref.~\cite{DOVER1984,Micaela2016,lopesPRD} to additional discussion) When we assume the SU(6) symmetry we have:

\begin{equation}
  \alpha_V = 1.00, \quad    z = \frac{1}{\sqrt{6}},  \quad \theta_V = 35.264, \label{SU6}
\end{equation}
and the Sakrurai proposals~\cite{SAKURAI1960} are restored: the $\rho$ meson couples to the isospin, therefore $ g_{\Sigma\Lambda\rho}$ = 0.

 As $\alpha_V$ = $F/(F+D)$  is a weight factor for the contributions of the
antisymmetric $F$ (corresponding to  $\{8'\}$) and the symmetric $D$ (corresponding to $\{8\}$) couplings relative to each other, when we assume $\alpha_V$ = 1, the symmetric couplings is neglect (g =0), therefore SU(3) CG coefficients already form a complete set without the need of the $g_{\Sigma^0\Lambda\rho}$ coupling. 

 However, if $\alpha_V~\neq~1$,  the symmetric coupling is also taken into account. Consequently the $ g_{\Sigma\Lambda\rho}~\neq~ 0$ and these interactions must be considered to account for the completeness of the theory. The now complete set of coupling constants in agreement with the SU(3) theory is presented in Tab.~\ref{Tc}. These results are fully model-independent and can be applied to a diversity of calculations in future works. { It is worth to point that $\alpha_V$ = 1 is still a legitimate choice and was used to reproduce hyperon-nucleon scattering data~\cite{Rijken2010}. The phenomenological necessity of the $g_{\Lambda^0\Sigma\rho}$ coupling in the context of the hyperon-nucleon scatterings remains unknown. 

\begin{table}[!t]
\begin{center}
\begin{tabular}{c|cccc}
&\multicolumn{4}{c}{$\alpha_v$}\\
  &   1.00&0.75&0.50&0.25   \\
\hline
 $g_{\Lambda\Lambda\omega}/g_{NN\omega}$        & 0.667 & 0.687   & 0.714 & 0.75   \\

 $g_{\Sigma\Sigma\omega}/g_{NN\omega}$         & 0.667 & 0.812  & 1.0 & 1.25   \\
 
$g_{\Xi\Xi\omega}/g_{NN\omega}$           & 0.333 & 0.437  & 0.571 & 0.75   \\
\hline
$g_{\Lambda\Lambda\phi}/g_{NN\omega}$  & -0.471 & -0.619  & -0.808 & -1.06   \\

$g_{\Sigma\Sigma\phi}/g_{NN\omega}$           & -0.471 & -0.441  & -0.404 & -0.354   \\

$g_{\Xi\Xi\phi}/g_{NN\omega}$           & -0.943 & -0.972  & -1.01 & -1.06   \\
\hline
$g_{\Lambda\Lambda\rho}/g_{NN\rho}$  &0.0&0.0&0.0&0.0\\
$g_{\Sigma\Sigma\rho}/g_{NN\rho}$           & 2.0 & 1.5  & 1.0 & 0.5   \\

$g_{\Xi\Xi\rho}/g_{NN\rho}$           & 1.0 & 0.5  & 0.0 & -0.5   \\
\hline
$g_{\Sigma^0\Lambda\rho}/g_{NN\rho}$           & 0.0 & 0.288  & 0.577 & 0.866   \\
\hline
\end{tabular}
 \caption{Complete set of baryon-vector mesons coupling constants for different values of $\alpha_v$, within the SU(3) symmetry group. These results are fully model-independent.}\label{Tc}
 \end{center}
 \end{table}

\section{The QHD Formalism and numerical results}

  I now study the effects of $g_{\Sigma^0\Lambda\rho}$ on dense nuclear matter  for $\alpha_V~\neq$ 1 and compare the results with those without this term.

I began by imposing  chemical equilibrium and zero electric charge net, a situation expected in neutron star interiors, to investigate the influence of the crossed terms. Let us start with a classical QHD Lagrangian without crossed couplings. Its Lagrangian  reads~\cite{Miyatsu2013,lopesPRD}:

\begin{eqnarray}
\mathcal{L} =  \sum_B\bar{\psi}_B[\gamma^\mu(\mbox{i}\partial_\mu  - g_{B\omega}\omega_\mu  -g_{B\phi}\phi_\mu   - g_{B\rho} \frac{1}{2}\vec{\tau} \cdot \vec{\rho}_\mu) \nonumber \\
- (M_B - g_{B\sigma}\sigma)]\psi_B  -U(\sigma) + \frac{1}{2}(\partial_\mu \sigma \partial^\mu \sigma - m_s^2\sigma^2)  \nonumber   \\
   - \frac{1}{4}\Omega^{\mu \nu}\Omega_{\mu \nu} + \frac{1}{2} m_v^2 \omega_\mu \omega^\mu+ \Lambda_{\omega\rho}(g_{\rho}^2 \vec{\rho^\mu} \cdot \vec{\rho_\mu}) (g_{\omega}^2 \omega^\mu \omega_\mu)  \nonumber \\
- \frac{1}{4}\Phi^{\mu \nu}\Phi_{\mu \nu} + \frac{1}{2} m_\phi^2 \phi_\mu \phi^\mu + \frac{1}{2} m_\rho^2 \vec{\rho}_\mu \cdot \vec{\rho}^{ \; \mu} - \frac{1}{4}\bf{P}^{\mu \nu} \cdot \bf{P}_{\mu \nu}  , \nonumber \\ \label{QHD} 
\end{eqnarray}
in natural units. Additional discussion about the parameters  and the formalism can be found in ref.~\cite{Miyatsu2013,lopesnpa,WALECKA1974,Serot_1992} and the references therein. The g's in Eq.~\ref{QHD} have only two instead three subscripts to let clear that in this Lagrangian $\bar{\psi}_B$ is always the complex conjugate of $\psi_B$. Applying Euler-Lagrange and the quantization rules we obtain the energy eigenvalues  (which at T = 0 K is also the chemical potential). In MFA we have:

\begin{equation}
E_B =  \sqrt{M^{*2}_B + k^2} + g_{B\omega}\omega_0 + g_{B\phi}\phi_0 + \frac{\tau_3}{2} g_{B\rho}\rho_0 \label{eigen}
\end{equation}

Now I add the coupled channels in the Lagrangian of Eq.~\ref{QHD}:

\begin{equation}
  \mathcal{L}_{\Lambda\Sigma^0\rho}  = -\frac{1}{2}g_{\Sigma\Lambda\rho}(\bar{\psi}_{\Lambda}\psi_\Sigma + \bar{\psi}_\Sigma\psi_\Lambda)\rho_0 , \label{crossL}
\end{equation}
where the $1/2$ factor was added to keep the internal coherence with Eq.~\ref{QHD}. When we apply Euler-Lagrange to now complete SU(3) Lagrangian, we see that the energy eigenvalue for all other six baryons is kept as in Eq.~\ref{eigen}. For the $\Lambda$ and the $\Sigma^0$ we have two coupled equations:

\begin{equation}
    \begin{cases}
    [\gamma^\mu(\mbox{i}\partial_\mu  - g_{\Lambda\omega}\omega_\mu) - M^{*}_\Lambda]\psi_\Lambda - \frac{1}{2}(g_{\Sigma^0\Lambda\rho})\rho_0\psi_\Sigma = 0 \\
     [\gamma^\mu(\mbox{i}\partial_\mu  - g_{\Sigma\omega}\omega_\mu) - M^{*}_\Sigma]\psi_\Sigma - \frac{1}{2}(g_{\Lambda\Sigma^0\rho})\rho_0\psi_\Lambda = 0 . \label{CE}
    \end{cases}
\end{equation}

 However, as we already know the energy eigenvalue without the coupled channel, their inclusion is much easier in Hamiltonian formalism.  The diagonal terms are the well-known unperturbed energy eigenvalues given by Eq.~\ref{eigen}, while the crossed terms are off-diagonal. We have:

 \begin{equation}
H = \left (
\begin{array}{cc}
   E_B  & \Delta \\
   \Delta  & E_B
\end{array}
\right )~\mbox{and}~H|\psi_B\rangle = E|\psi_B \rangle , \label{hamilton}
 \end{equation}
 where $|\psi_B \rangle  = (\psi_\Lambda, \psi_\Sigma)$ and $\Delta =  1/2(g_{\Sigma^0\Lambda\rho})\rho_0$. As we are dealing with a beta-stable matter, $\mu_\Lambda = \mu_\Sigma$, the new energy eigenvalues are (see for instance chapter 5 of Sakurai's book~\cite{Sakuraibook} for a complete discussion):

 \begin{equation}
 \begin{array}{c}
      E_1 = \sqrt{M_\Lambda^{*2} + k^2}  + g_{\Lambda\omega}\omega_0  + g_{\Lambda\phi}\phi_0 -\frac{g_{\Sigma\Lambda\rho}}{2}\rho_0,\\
    E_2 = \sqrt{M_\Sigma^{*2} + k^2}  + g_{\Sigma\omega}\omega_0  + g_{\Sigma\phi}\phi_0 + \frac{g_{\Sigma\Lambda\rho}}{2}\rho_0.
 \end{array} \label{E12}
 \end{equation}

 Despite  the energy eigenvalues from Eq.~\ref{E12} being exact, the issue here is that the coupled channels lead us to mixed states~\cite{PROVIDENCIA1985}. In other words, the $\psi_\Lambda$ and $\psi_\Sigma$ are not eigenstates of the Hamiltonian of Eq.~\ref{hamilton} anymore. Instead, we have a  superposition~\cite{Sakuraibook,PROVIDENCIA1985}.  However, as we have $E_B >> \Delta $ in Eq.~\ref{hamilton}, and following Sakurai's nomenclature~\cite{Sakuraibook}, $\psi_\Lambda$ and $\psi_\Sigma$ are ''almost good" eigenstates of Eq.~\ref{hamilton}. Therefore we can recognize $E_1$ as the eigenvalue of the $\Lambda$ and $E_2$ as the eigenvalue of the $\Sigma^0$. 
 
Reducing the coupled channel to MFA is not new. It was successfully used to account for the kaon interaction in nuclear medium in MFA (see, for instance, section 10.1 of Glendenning's book~\cite{Glenbook} and the references therein.), though such interaction is explicitly a coupled channel coming from the $g_{N\Lambda K}$ and $g_{N\Sigma K}$ couplings ~\cite{Weise1988,KOCHPLB1994} (indeed, as the $g_{\Lambda\Lambda\rho}$, the $g_{NNK}$ is null~\cite{Swart1963}). It is also worth to point that the $\Lambda$-$\Sigma$ interaction is supported by experimental data, in the so-called coherent $\Lambda-\Sigma$ coupling~\cite{Aka2000,Aka2002} Finally, the eigenvalues of the other six baryons are given by their usual expression, Eq.~\ref{eigen}.

 It is interesting to notice that when I calculated the $g_{\Sigma^0\Lambda\rho}$ and the $g_{\Lambda\Sigma^0\rho}$
 coupling constants from the SU(3) Clebsch-Gordan coefficients, I showed that both have positive signs. However, as they are off-diagonal contributions, they ultimately contribute with opposite signs to the energy eigenvalues, as displayed in Eq.~\ref{E12}. So, for practical purposes, the ($\Sigma^0,~\Lambda$) forms a new isospin doubled, exactly as the (p,n), ($\Sigma^+,~\Sigma^-$), and ($\Xi^0,~\Xi^-$), with the coupling constants given by Tab.~\ref{Tc}.
 The total EoS is given by~\cite{Miyatsu2013}:

\begin{eqnarray}
 \epsilon = \sum_B \frac{1}{\pi^2}\int_0^{k_{Bf}} dk k^2 \sqrt{k^2 + M_B^{*2}} +U(\sigma_0)  
 +\frac{1}{2}m_\sigma^2\sigma_0^2 + \frac{1}{2}m_\omega^2\omega_0^2 \nonumber \\ 
 + \frac{1}{2}m_\phi^2\phi_0^2 + \frac{1}{2}m_\rho^2\rho_0^2 
  +3 \Lambda_v\omega_0^2\rho_0^2 
  \label{EL7}  + \sum_l \frac{1}{\pi^2}\int_0^{k_{lf}} dk k^2 \sqrt{k^2 + m_l^{2}} , 
\end{eqnarray}
where $B$ indicats baryon and $l$ indicates leptons. The pressure is easily obtained by thermodynamic relations: $p =
\sum_f \mu_f n_f - \epsilon$, where the sum runs over all the
fermions and $\mu_f$ is the corresponding chemical potential.

  To obtain numerical results, I  consider $\alpha_V$ a free paramenter but  use only $\alpha_V$ = 0.25, which has the strongest influence of the $g_{\Sigma^0\Lambda\rho}$, in order to not saturate the figures. Also, I use two different parameterizations, the eL3$\omega\rho$~\cite{lopesPRD}, that virtually fulfill every constraint of the symmetric nuclear matter, and the well-known and the widely used GM1 paramertrization~\cite{GM1}.  All parameters and predictions for the eL3$\omega\rho$ are presented in Tab. I of ref.~\cite{lopesPRD}, while the GM1 can be found in Tab. I of ref.~\cite{lopesnpa}. The coupling constants of the hyperons with the scalar meson are fixed to reproduce the hyperon potential depth values: $U_\Lambda$ = - 28 MeV and $U_\Sigma$ = + 30 MeV. For the   and $U_\Xi$, I chose  $U_\Xi$ = -18 MeV as suggested in ref.~\cite{Potentials} when I use the GM1 parametrization (which allows a direct comparison with the results presented in ref.~\cite{lopesnpa}),  and chose $U_\Xi$ = - 4 MeV as suggested in ref.~\cite{LQCD} for the eL3$\omega\rho$ parametrization (which allow us a comparison with the results presented in ref.~\cite{lopesPRD} ). 

The reason I use two different parametrizations is that in the eL3$\omega\rho$ there is a non-linear coupling between the $\omega$ and $\rho$ mesons, as introduced in the IUFSU model~\cite{IUFSU}, while for the GM1 there isn't.   Such coupling influences the mass of the $\rho$ meson, which ultimately affects the strength of the $\rho$ field at high densities.

\begin{figure*}[t]
\begin{tabular}{cc}
\centering 
\includegraphics[scale=.51, angle=270]{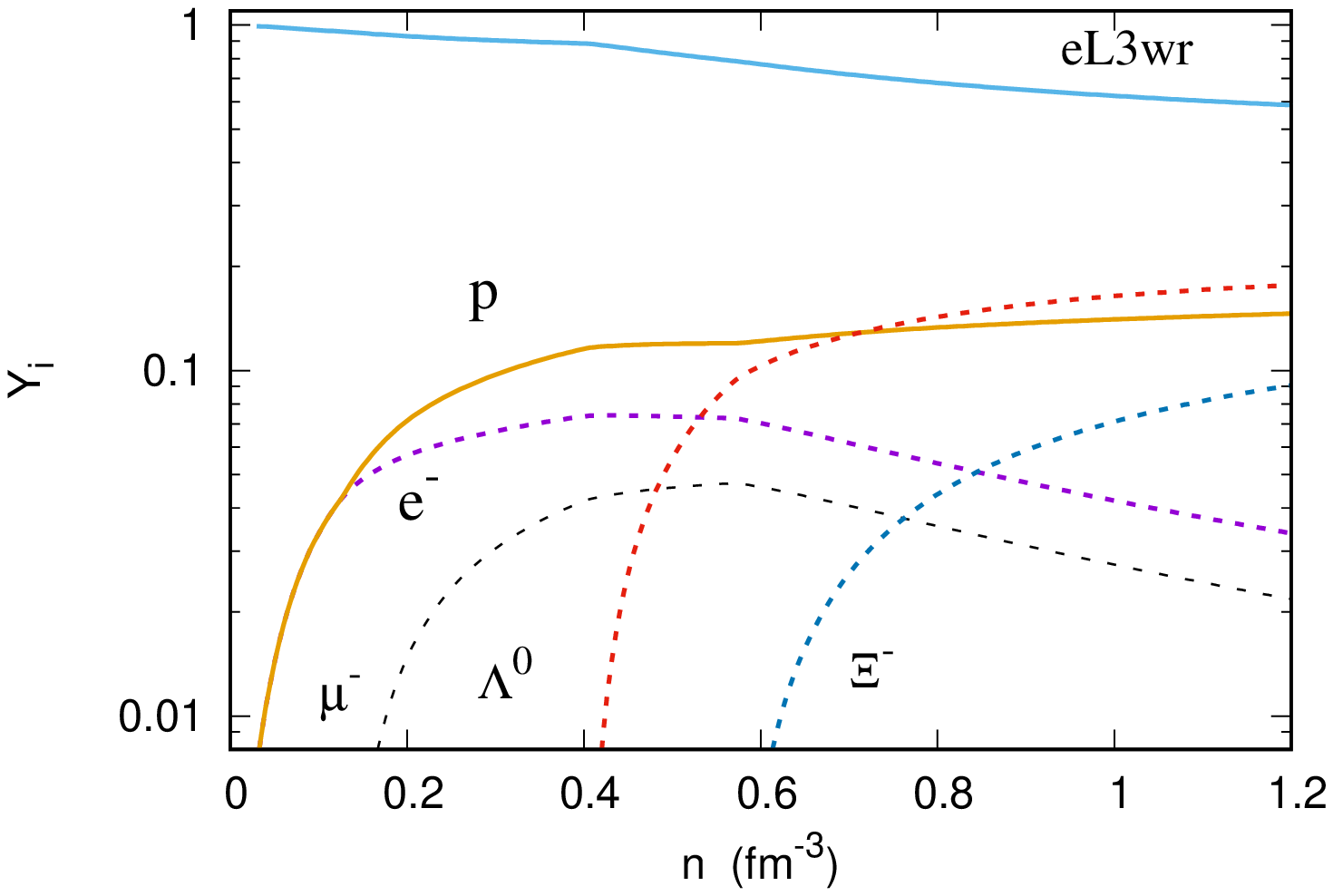} &
\includegraphics[scale=.51, angle=270]{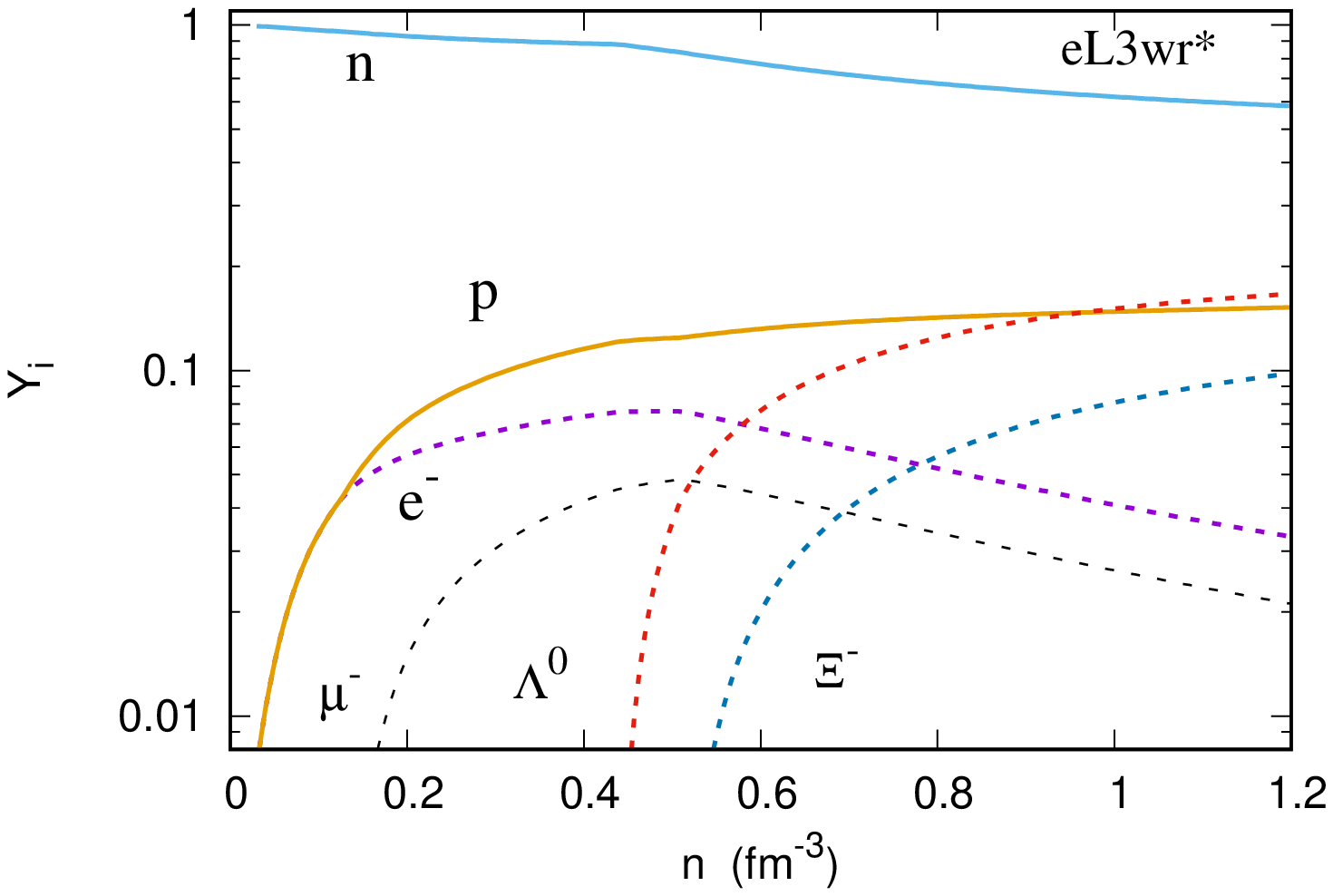}\\
\includegraphics[scale=.51, angle=270]{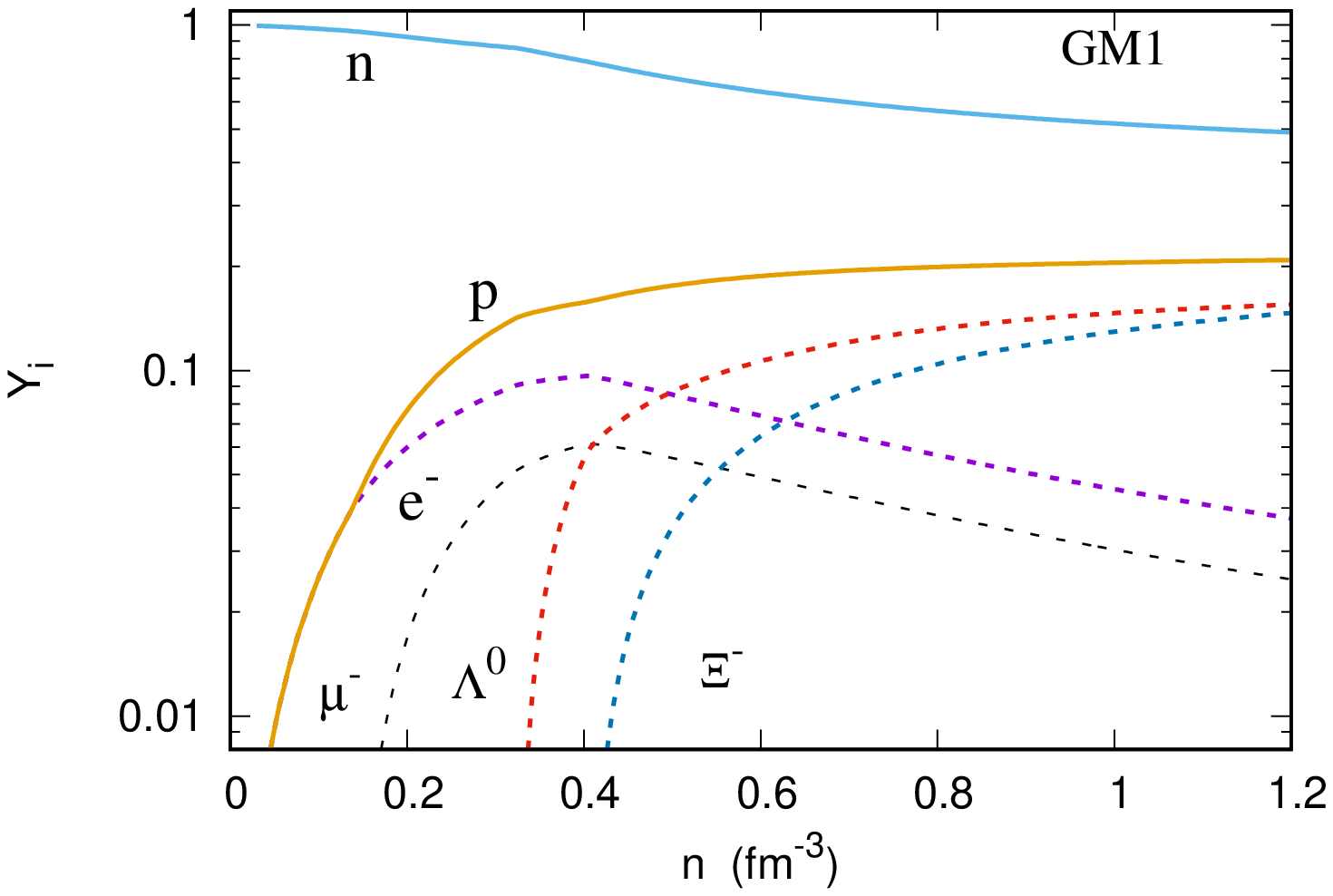} &
\includegraphics[scale=.51, angle=270]{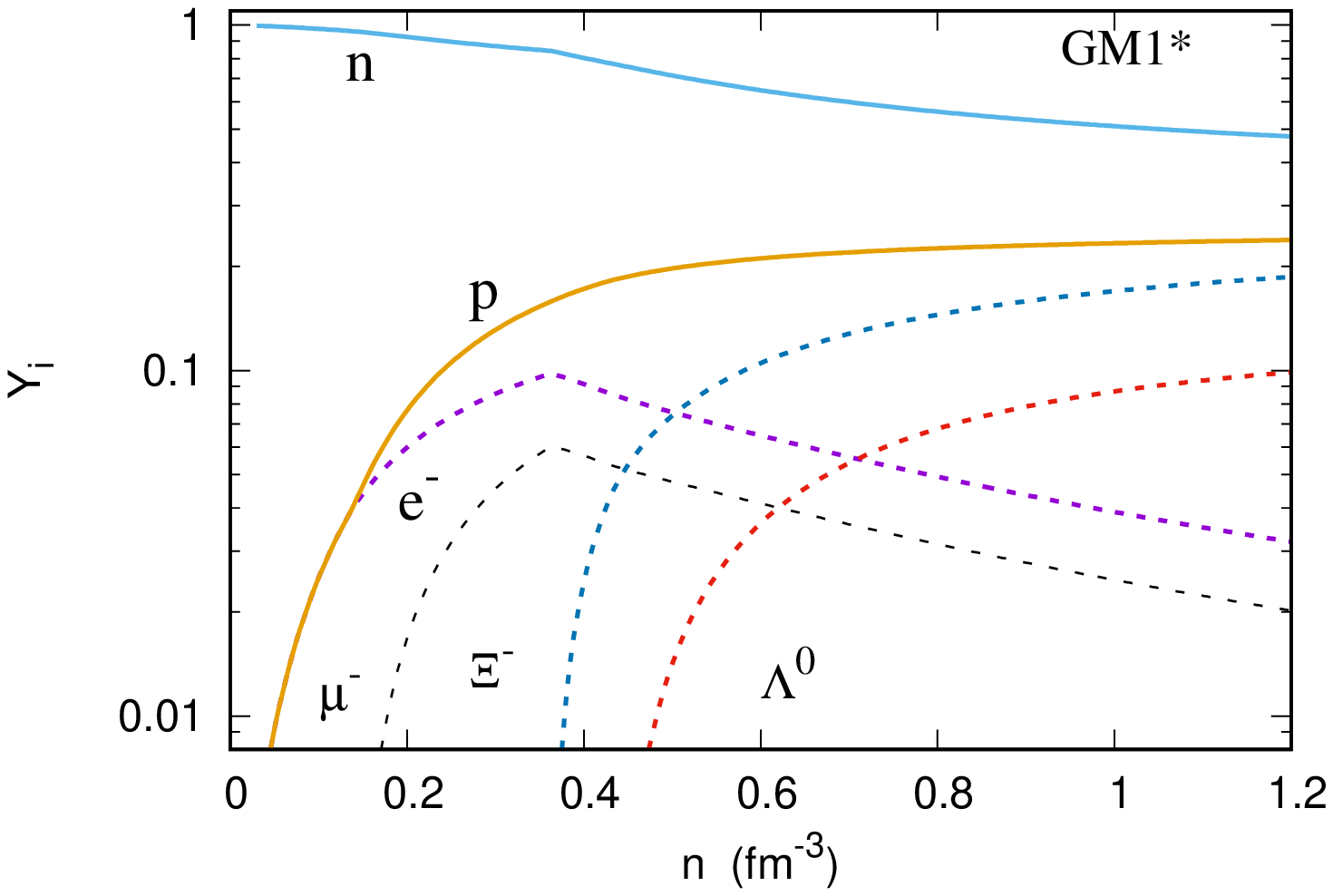}\\
\end{tabular}
\caption{(Color online) Particle population for the eL$\omega\rho$ and for the GM1. Results with (without) * indicate the presence (absence) of the  $g_{\Sigma^0\Lambda\rho}$ coupling. } \label{ff}
\end{figure*}

 The particle population for the beta-stable matter at T = 0 K for $\alpha_v$ = 0.25 is displayed in Fig.~\ref{ff}. We can see that the main effect of the $g_{\Sigma^0\Lambda\rho}$ coupling is to suppress the $\Lambda$ onset, pushing it away to higher densities, whilst, at the same time, it favors the $\Xi^-$.
 In the case of the eL3$\omega\rho$ parametrization, the prensece of the $g_{\Sigma^0\Lambda\rho}$ coupling,
 pushes the $\Lambda$ threshold from 0.4114 fm$^{-3}$ to 0.4416 fm$^{-3}$, whilst the $\Xi^-$ is draw close-approach from from 0.5821 fm$^{-3}$ to 0.5168 fm$^{-3}$. This indicates an increase of around $10\%$ in the density of the $\Lambda$ and a decrease of around $10\%$ in the density of the $\Xi^-$.
 In the case of the GM1 parametrizations, the results are more extreme. The  $g_{\Sigma^0\Lambda\rho}$ coupling not only suppresses the $\Lambda$ threshold whilst favoring the $\Xi^-$, but it exchanges their roles.
 Within it, the $\Xi^-$ is now the first hyperon to appear and becomes the most populous hyperon at higher densities. The $\Lambda$ threshold is pushed away from 0.3264 fm$^{-3}$ to 0.4405 fm$^{-3}$; an increae of around $35\%$. On the other hand, the $\Xi^-$ is drawn close-approach from from 0.4079 fm$^{-3}$ to 0.3655 fm$^{-3}$, a decrease of around $10\%$.

Now I use the EoS of the beta-stable electric neutral matter to solve the TOV equations~\cite{TOV} equations. For both parametrizations, I use the BPS EoS~\cite{BPS} for the outer crust and the BBP EoS~\cite{BBP} for the inner crust. I do not plot the EoS itself because the effects of the $g_{\Sigma^0\Lambda\rho}$ coupling are visually indistinguishable.  The numerical results are presented in  Fig.~\ref{F1}. 

We can also discuss some constraints related to neutron stars. Today, maybe the more important constraint is the undoubted existence of supermassive neutron stars. Using the NICER x-ray telescope, ref.~\cite{Riley2021} was able to constraint the mass and the radius of the PSR J0740 + 6620 in the range of $M = 2.08 \pm 0.07 M_\odot$, and 11.41 km $<~R~<$ 13.70 km respectively. We plot this constraint as a hatched area in Fig.~\ref{F1}. As can be seen, both the eL3$\omega\rho$ and the GM1 fulfill this constraint.

Other constraints are related to the radius and tidal parameter of the canonical 1.4 $M_\odot$ star, however, they are still the subject of high debate about their true values.  Recently,  results obtained from Bayesian analysis indicate that the radius of the canonical star lies between 
10.8 km and 13.2 km ~\cite{Yuxi}; and 11.3 km to 13.5 km ~\cite{Michael};
whilst results coming from the NICER x-ray telescope points out that  $R_{1.4}$ lies between
11.52 km and 13.85 km from ref.~\cite{Riley:2019yda} and between 11.96 km and 14.26 km from ref. ~\cite{Miller:2019cac}. State-of-the-art theoretical results at low and high baryon density point to an upper limit of $R_{1.4}$ $<$ 13.6 km~\citep{Annala2}. Finally, PREX2 results~\citep{PREX2} indicate that the radius of the canonical star lies between 13.25 km $<~R_{1.4}~<$ 14.26 km. I use the intersection between the two NICER results~\cite{Miller:2019cac,Riley:2019yda}: 11.96 km $<~R_{1.4}~<$ 13.85 km as a constraint for the canonical star.

\begin{figure}[ht]
  \begin{centering}
\begin{tabular}{c}
\includegraphics[width=0.35\textwidth,angle=270]{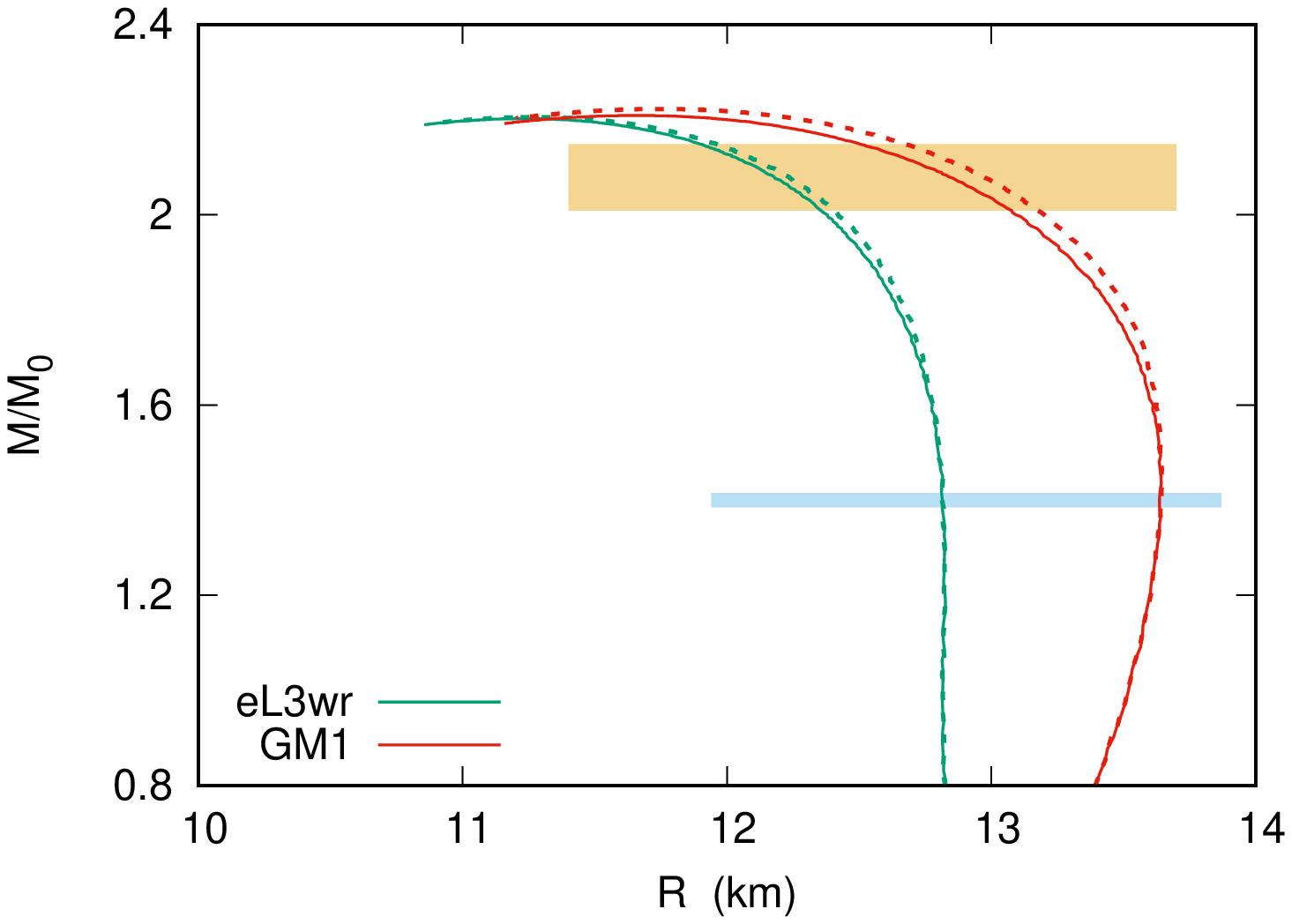} 
\includegraphics[width=0.35\textwidth,angle=270]{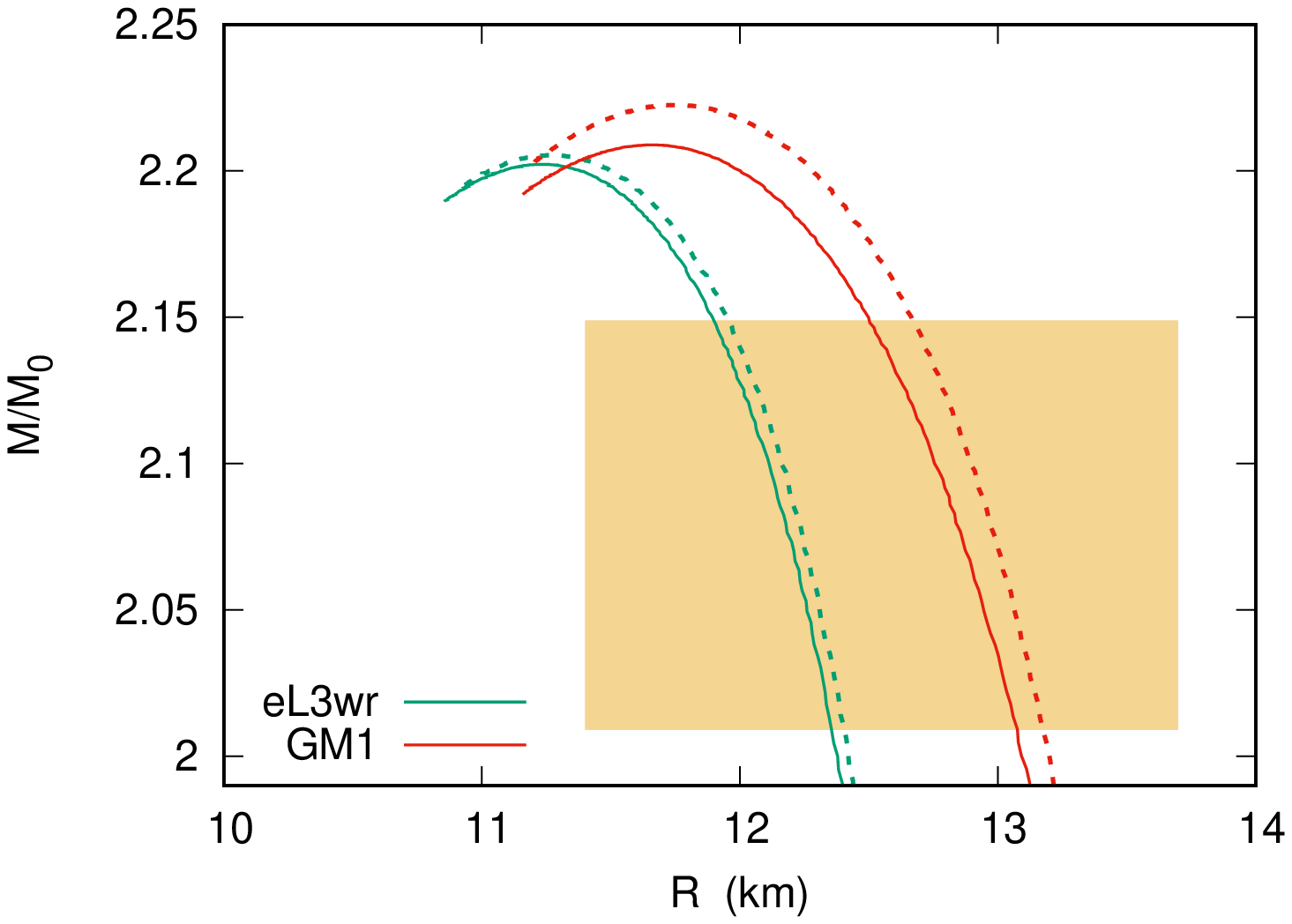} \\
\end{tabular}
\caption{(Color online) Left: Neutron stars mass-radius relation for the eL$\omega\rho$ and the GM1 models. The solid (dotted) lines indicate the presence (absence) of the $g_{\Sigma^0\Lambda\rho}$ coupling. The orangey hatched area is the mass-radius uncertainty of the PSR J0740+6620 pulsar~\cite{Riley2021}, and the bluish hatched area is the intersection of two estimative  from  NICER  for the 1.4$M_\odot$~\cite{Riley:2019yda,Miller:2019cac}. Right: Zoom in for $M~\geq~2.0 M_\odot$.} \label{F1}
\end{centering}
\end{figure}

In relation to the tidal parameter, an upper limit of 860 was found in ref.
~\cite{Michael}. A  close limit, $\Lambda_{1.4}~<$ 800 was pointed out in ref. ~\cite{tidal4}. In ref. ~\cite{Yuxi}, an upper limit of 686 was deduced from  Bayesian analysis. On the other hand,
two mutually exclusive constraints are presented in ref.
~\cite{tidal1}, which proposed a limit between 70 $<~\Lambda_{1.4}~<$ 580, and  the PREX2 inferred values, whose  limit lies between 642 $<~\Lambda_{1.4}~<$ 955~\citep{PREX2}.  As hyperons are not present at a 1.4 $M_\odot$ star, we always have $R_1.4$ = 12.82 km and $\Lambda_{1.4}$ = 516 for the eL3$\omega\rho$, and 
$R_{1.4}$ = 13.68 km and $\Lambda_{1.4}$ = 696 for the GM1. Other results are presented in Tab.~\ref{T2}.

\begin{center}
\begin{table}[ht]
\begin{center}
\begin{tabular}{|c|c|c|c|c|}
\hline 
   & $M_{max}/M_\odot$ & Hyp.  at (fm$^{-3}$) & $R_{2.0}$  (km) \\
 \hline
 eL3$\omega\rho$ & 2.202 & $\Lambda$ at 0.4114   &12.379 \\
 eL3$\omega\rho$* & 2.206 & $\Lambda$ at 0.4416  &12.420 \\
  \hline
 GM1& 2.223 & $\Lambda$ at 0.3264  &13.092   \\
GM1* &  2.208  & $\Xi^-$ at  0.3655   &13.193  \\
\hline
\end{tabular}
\caption{ Some of the neutron stars properties. Results with (without) * indicate the presence (absence) of the $g_{\Sigma^0\Lambda\rho}$ coupling.} 
\label{T2}
\end{center}
\end{table}
\end{center}

As can be seen, for massive neutron stars the influence of the $g_{\Sigma^0\Lambda\rho}$ coupling is very limited. The $g_{\Sigma^0\Lambda\rho}$ coupling causes a small increase of the maximum mass, as well causes an increase of the radius for a fixed mass value. All these increments are about only 0.5$\%$. This may sound a little disappointing but we must remember that no one could know how strong would be the influence of the $g_{\Sigma^0\Lambda\rho}$ until someone calculated its value.

\begin{figure}[h!]
  \begin{center}
\begin{tabular}{c}
\includegraphics[width=0.333\textwidth,angle=270]{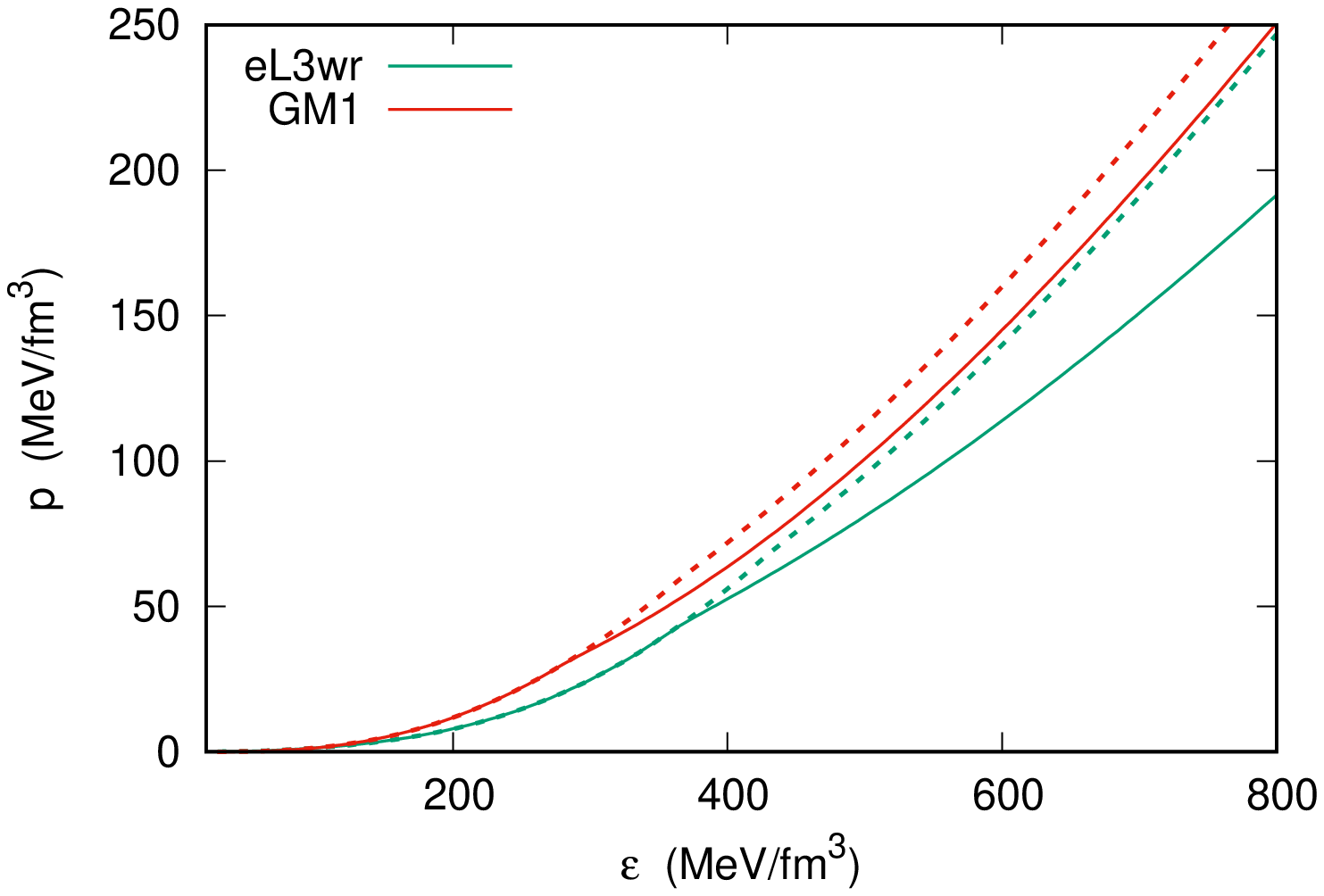} 
\includegraphics[width=0.333\textwidth,,angle=270]{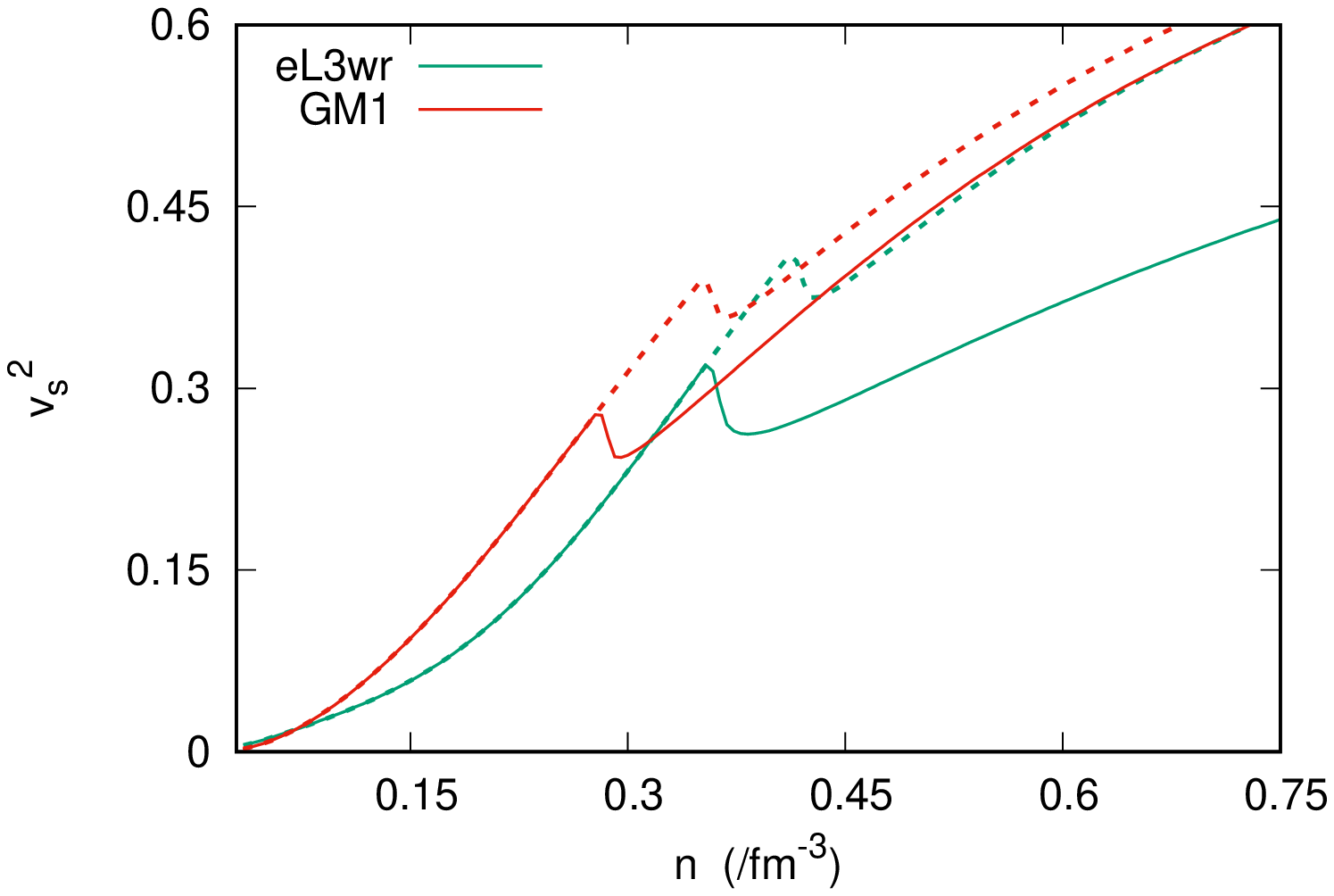} 
\end{tabular}
\caption{(Color online) The EoS and the $v_s^2$ for neutron-$\Lambda$ matter. The dotted (solid) lines indicate the presence (absence) of the $g_{\Sigma^0\Lambda\rho}$ coupling.} \label{F2}
\end{center}
\end{figure}

The effect of the $g_{\Sigma^0\omega\rho}$ coupling is more evident when we consider a matter consisting of only neutrons and $\Lambda$'s. In ref.~\cite{Gul2012,James2016} the authors study a liquid-gas-like phase transition within neutron-$\Lambda$ matter. The neutron-$\Lambda$ matter was also used to study spinodal instability in ref.~\cite{James2017}. Moreover, the existence of a neutral bound state consisting of only neutrons and $\Lambda$'s was investigated in ref.~\cite{Rappold2013,Garcilazo2017}. Here I follow ref.~\cite{James2017} and use $\mu_n =\mu_\Lambda$. The EoS and the square of the speed of sound $v_s^2 = \partial p/\partial \epsilon$ are displayed in Fig.~\ref{F2}.

As can be seen, the presence of the $g_{\Sigma^0\Lambda\rho}$ stiffens the EoS, as well as increases the speed of sound at high densities and pushes away the onset of the $\Lambda$. For the eL3$\omega\rho$ the $\Lambda$ threshold is pushed from 0.3634 fm$^{-3}$ to 0.4164 fm$^{-3}$, while within the GM1 parametrization the onset is pushed from 0.2819 fm$^{-3}$ to 0.3586 fm$^{-3}$.  For the GM1 the increase of the density threshold is higher than 25$\%$, while for the  eL3$\omega\rho$ it is around $15\%$.

 Before I finish, I would like to mention that the applications of the $g_{\Sigma^0\Lambda\rho}$
are far beyond those presented in this work. For instance, it can potentially affect  hypernuclei~\cite{Aka2000} energy levels, as well as hyperon-baryon scattering~\cite{Rijken2010}. 
\vspace{2.0 cm}

\section{Conclusions}

In this work, I investigate the use of the symmetry groups and the SU(3) Clebsch-Gordan coefficients to fix the coupling constants of the baryon octet   with the vector meson in order to keep the Yukawa Lagrangian as a singlet for both: the antisymmetric and symmetric couplings. The main results of the present work are summarized below:

\begin{itemize}

    \item 
I found that the current set of coupling constants for the SU(3) symmetry group does not satisfy the relations of completeness and closure for the symmetric coupling, while was already complete for the antisymmetric one ($\alpha_V$ = 1).

\item
There are two  additional Yukawa interactions related to the exchange of the neutral $\rho$ meson between the $\Sigma^0$ and the $\Lambda$ hyperon. When these interactions are taken into account the relations of completeness and closure are restored.

\item 
Then I calculate the $g_{\Sigma^0\Lambda\rho}$ coupling constants within SU(3) and SU(6) symmetry groups.
In SU(6) we have $g_{\Sigma^0\Lambda\rho}$ = 0, and Sakurai's theory of strong interaction is restored~\cite{SAKURAI1960}. Therefore, for the pure $F$-mode ($\alpha_V$ = 1) the $g_{\Sigma^0\Lambda\rho}$ is not required to satisfy the SU(3) symmetry group. However, if $\alpha_V~\neq~1$, the $ g_{\Sigma\Lambda\rho}~\neq~ 0$ and these interactions must be considered to account for the completeness of the theory. These results are fully model-independent.

\item 
The $\Lambda$-$\Sigma$ interaction is supported by experimental data, in the so-called coherent $\Lambda-\Sigma$ coupling~\cite{Aka2000,Aka2002}

In order to study the effects of the $g_{\Sigma^0\Lambda\rho}$ couplings, I add these crossed  Yukawa couplings to the QHD model  to study dense nuclear matter, on which $\alpha_V$ is usually a free parameter.

\item 
I show that these crossed terms enter as off-diagonal terms in the Hamiltonian. As a consequence, the coupling with the $\Lambda$ and with the $\Sigma^0$ present opposite signs, despite having the same Clebsch-Gordan coefficients.

\item
I then obtain some numerical results  for dense nuclear matter  within two different parametrizations: the eL$\omega\rho$~\cite{lopesPRD} and the GM1~\cite{GM1}. I show that the $g_{\Sigma^0\Lambda\rho}$ coupling
suppresses the $\Lambda$ onset whilst favoring the $\Xi^-$ one. In the case of the GM1, this is enough to 
make the $\Xi^-$ the first hyperon to appear. In the case of massive neutron stars,  the $g_{\Sigma^0\Lambda\rho}$ coupling causes a very small increase of the maximum masses and the radii for fixed masses (around $0.5\%$).

\item
Finally, I study a hadronic matter constituted by only neutrons and $\Lambda$'s. I show that the $g_{\Sigma^0\Lambda\rho}$ coupling stiffens the EoS, and pushes the hyperon threshold to higher densities. It also affects the speed of the sound.

\end{itemize}

\section*{Acknowledgements}

The author was partially supported by CNPq Universal Grant No. 409029/2021-1.

\bibliographystyle{ptephy}
\bibliography{sample}
%

\vspace{0.2cm}
\noindent


\let\doi\relax


\

\end{document}